\documentclass[english]{elsart}
\usepackage{graphicx}
\usepackage{babel}
\usepackage[T1]{fontenc}

\begin{document}

\begin{frontmatter}
\title{An Extra Push from Entrance-Channel Effects}

\author[ires]{N. Rowley}
\corauth[cor]{Corresponding author}
\ead{neil.rowley@ires.in2p3.fr}
\author[use]{N. Grar}
\author[tus]{K. Hagino}

\address[ires]{Institut de Recherches Subatomiques/Universit\'e Louis Pasteur 
(UMR 7500), 23~rue du Loess, F-67037 Strasbourg, France}
\address[use]{DAC, Department of Physics, Universit\'e de S\'etif, Alg\'erie}
\address[tus]{Department of Physics, Tohoku University, Sendai
980-8578, Japan}
\date{\today}

\begin{abstract}
Symmetric heavy-ion collisions are known to display an `extra-push' effect.
That is, the energy at which the s-wave transmission is 0.5
lies significantly higher than the nominal Coulomb barrier. Despite this,
however, the capture cross section is still greatly enhanced below the
uncoupled barrier. It is shown that this phenomenon can be simply explained
in terms of entrance-channel effects which account for long-range Coulomb
excitations.
\end{abstract}

\end{frontmatter}

When reactions between intermediate-mass heavy ions lead to non-fissile
composite systems,
the relationship between the cross sections for capture (passing over 
or penetrating through the
Coulomb barrier), fusion (evolution to a compact equilibrated compound
nucleus; CN) and evaporation residues (ER) is straightforward.
If fission is unimportant, all of the
above cross sections are essentially equal: $\sigma_{\rm cap}=
\sigma_{\rm fus}=\sigma_{\rm ER}$.
Of course it is well 
known that couplings to collective states of the target and projectile
can lead to a distribution of Coulomb barriers~\cite{review} but this
does not in any way change the above relationship, any structure
in $\sigma_{\rm cap}$ also being present in $\sigma_{\rm ER}$. To study the 
effects of the entrance channel, one may simply measure the 
long-lived evaporation residues which recoil in a relatively narrow 
cone around the beam direction 
(dispersed by the emission of neutrons, protons and $\alpha$-particles 
from the CN). The results for intermediate-mass systems almost
invariably show that collective couplings increase the sub-barrier capture
cross section (see, for example, Ref.~\cite{review}).

For heavier systems, other reactions mechanisms intervene and
complicate the situation both experimentally and theoretically. 
In particular, the composite system might not fuse but instead
quickly separate into two fragments similar in mass 
and charge to the target and projectile (quasifission; QF).
The CN itself may also fission (fusion-fission; FF) rather than 
decaying to a long-lived residue through particle evaporation. 
For very heavy systems the fission modes dominate and a 
complete understanding of the interplay between the various reaction
mechanisms is especially important in heavy-element creation.

To measure $\sigma_{\rm cap}$ directly in the general case,
$\sigma_{\rm ER}$, $\sigma_{\rm QF}$ and 
$\sigma_{\rm FF}$ must all be measured (including the fragment angular
distributions) in order to obtain $\sigma_{\rm cap}$. Though if 
quasifission is not thought to
be important, one could still try to obtain the capture cross section
by measuring only the evaporation residues, and using an evaporation-model
code that accounts for the competition between fusion-fission and
fusion-evaporation decay modes
to reconstruct the capture cross section required to
reproduce $\sigma_{\rm ER}$. This was the aim of a series of
experiments performed at GSI using projectiles and targets around
mass 100~\cite{quint,keller,othergsi1,othergsi2}. The interesting result
is that the capture cross sections obtained displayed 
a so-called extra-push effect. That is, the energy $\bar B$ at which the 
deduced s-wave transmission $T_0$ was 0.5, could greatly exceed the 
barrier height predicted by potential models such as that of Bass 
\cite{bass}. This in itself might be explained by an internal barrier
which must be crossed after passing the outer Coulomb barrier if fusion
is to take place, and this could be thought of as the conditional saddle point
in the liquid-drop nuclear potential~\cite{extrapush}. However, the data are
not entirely consistent with such a description since, despite the shift 
of the $T_0=0.5$ point to higher energies, $\sigma_{\rm cap}$ was 
still found, as for lighter systems, to be 
strongly enhanced at energies well below 
the Bass barrier. This enhancement was quantified by defining a single
(adiabatic) barrier $B_{\rm ad}$ which yielded the correct cross section 
at the very lowest energies, and thus obtaining an overall width of 
the barrier distribution $D_{\infty}=\bar B -B_{\rm ad}$. For the
system $^{100}$Mo + $^{100}$Mo, for example, it was found 
that $D_{\infty}\approx$ 20 MeV.

The authors of Ref.~\cite{quint} tried to
fit their data with an entrance-channel model using the 
simplified coupled-channels code
CCFUS~\cite{ccfus} with couplings to the known quadrupole- and octupole-phonon
states of target and projectile. They found that in general such 
calculations could account for only about one half of $D_{\infty}$.
The main aim of the present paper is to show that more complete
coupled-channels calculations are in fact capable of fitting $D_{\infty}$
rather well, and also yielding the correct shape of the capture cross section
(assumed by Quint et al. to arise from a gaussian barrier distribution;
see Fig. 2). 
An important ingredient missing from the earlier calculations will
be shown to be the long-range Coulomb couplings which polarise the 
target and projectile well before the Coulomb barrier is reached. The role
of multi-phonon excitations is also important.

The points in Fig.~1 shows both on a logarithmic scale 
and a linear scale the deduced experimental
s-wave transmission as a function of the incident energy $E_{\rm cm}$
for the system  $^{100}$Mo + $^{100}$Mo. They were 
derived by assuming a gaussian barrier distribution with a centroid 
$B$ and standard deviation $\Delta$ and varying these parameters until 
the fusion-evaporation-model code HIVAP~\cite{hivap} reproduced the 
evaporation-residue cross section. The experimental values of $T_0$ are 
then obtained through 
\begin{equation}
T_0^{\rm exp}=T_0^{\rm theory}\frac{\sigma_{\rm ER}^{\rm exp}}
{\sigma_{\rm ER}^{\rm HIVAP}}.
\label{cross}
\end{equation}
This is a very good way to 
represent the data, since the quantity $T_0$ is directly related to
the entrance-channel dynamics. 
However it should be stressed that the experimental $T_0$ are not true
experimental data. They depend not only on $B$ and $\Delta$ but also
on the parameters entering into the HIVAP calculation.
This leads to certain ambiguities for some system, a point to which we shall
return later. For the moment we accept these numbers at face value and
attempt to fit them with calculations using the program CCFULL~\cite{ccfull},
again using known phonon states in $^{100}$Mo. 

This nucleus has strong quadrupole- and octupole phonon states lying
at relatively low excitation energies and we shall use
the adopted empirical values of these energies and the corresponding 
deformation parameters: $E(2^+)=0.536$ MeV, $\beta_2$=0.21;
$E(3^-)=1.908$ MeV, $\beta_3$=0.17~\cite{defparas}. The only other parameters
entering our calculations are the no-coupling barrier height $B_{\rm nc}$,
which we shall vary to fit the data, and the diffusivity of the nuclear
potential for which we take a standard vaue of $a=0.6$ fm.

\begin{figure}
\vspace{100mm}
\includegraphics{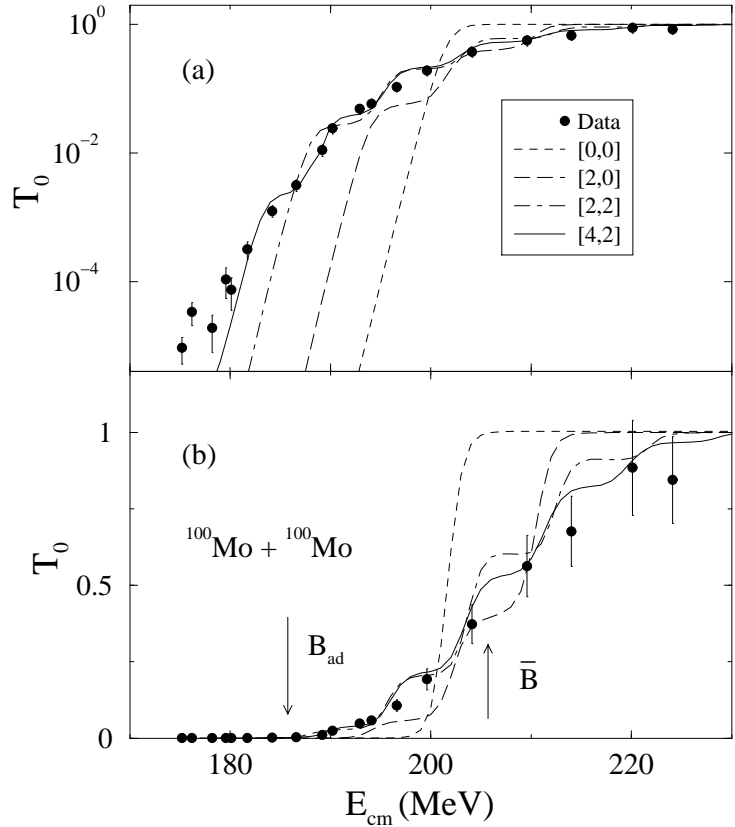}
\caption{Experimental $T_0$ compared with 
various CCFULL calculations with different numbers of
phonon excitations. See text for details. Arrows indicate
the average barrier $\bar B$ and the adiabatic barrier $B_{\rm ad}$,
whose difference gives $D_\infty$. Parts (a) and (b) show same curves but
on logarithmic and linear scales.
}
\label{fig1}
\end{figure}

The dashed curves in Fig.~1 a,b show the no-coupling result, which is
seen to greatly underestimate $T_0$ at low energies. The other curves show
calculations including various phonon couplings $[N_{\rm quad},N_{\rm oct}]$. 
The symmetry of the present system allows us to use a simple theoretical
trick to reduce the number of channels in a given calculation. For example, 
the calculation with one quadrupole phonon in both target and projectile,
along with the mutual excitation can be exactly treated as a two-channel
calculation with renormalised couplings. The details of this method will
be presented elsewhere~\cite{graretal}. Thus the calculation labelled
$[4,2]$ means two quadrupole- phonon excitations and one octupole excitation
in each nucleus along with all possible mutual excitations. It is 
clearly seen that as the complexity of the coupling increases, the 
theoretical results converge to the experimental curve both at high energies
(see linear scale) and low energies (logarithmic scale). The final
calculation $[4,2]$, however, still slightly underpredicts $T_0$ at
the very lowest energies, and it might be asked why we do not pursue 
this with a $[4,4]$ calculation.

The problem here is that the full coupled-channels calculations become
numerically unstable at low energies if too many channels
are included. The reason is that we are essentially
integrating the Schroedinger equation at energies around 30 MeV under
the highest effective barrier, and the energies losses due to couplings
to the phonon states further reduce the kinetic energy of the relative
motion. This problem increases with the number of phonon channels and
the program breaks down at the lowest energies. However, the problem 
may be overcome to some extent by reducing the width of the Coulomb barrier,
and this can be achieved by decreasing the diffusivity $a$. In Fig.~2
we show the results of calculations using $a=0.2$ fm. We should stress 
that we do not believe such a low value of the diffusivity but only use
it as a means of seeing the effect of the higher phonon couplings in the 
$[4,4]$ calculation. However, the use of $a=0.2$ changes rather little the
barrier positions. Its main effect is to decrease the rate at which the
cross section falls off below the Coulomb barrier. But since the 
cross section at low energies is dominated by the lowest barriers, this 
effect is only significant below the very lowest (adiabatic) barrier. 

\begin{figure}
\vspace{120mm}
\includegraphics{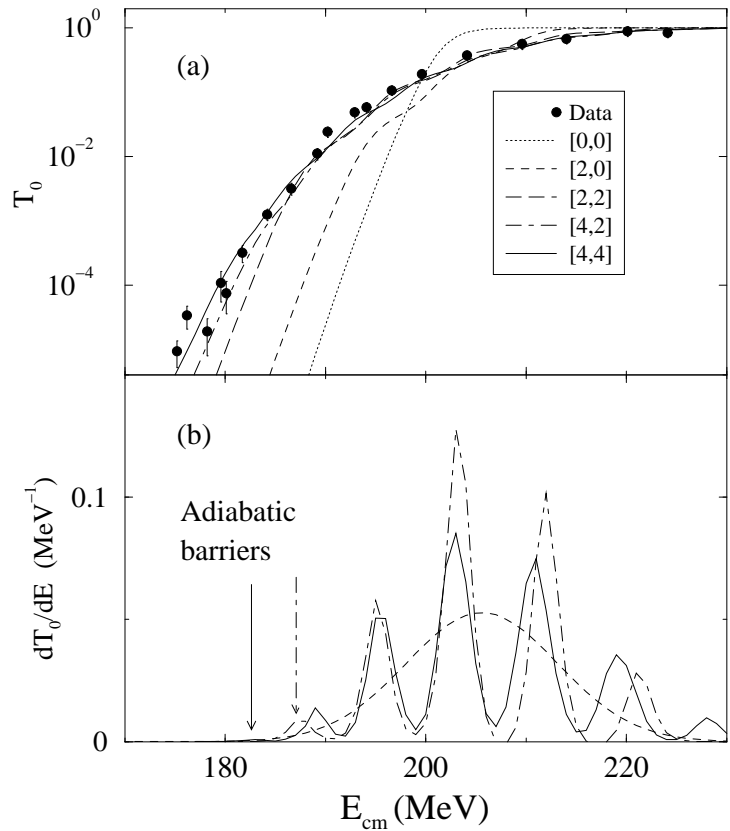}
\caption{Using $a=0.2$ fm permits the $[4,4]$ calculation with both 
double quadrupole- and octupole-phonon excitations. Note that the 
calculations have virtually converged, with a new lowest barrier 
emerging but with very small weight. The dashed curve in (b) is the 
gaussian barrier distribution of Ref.~\cite{quint}.
}
\label{fig2}
\end{figure}

We show again in Fig. 2a the calculations with the same coupling schemes
as in Fig.~1, and note that the inclusion of the double-octupole phonon
shifts the low-energy cross section down by about a further 2 MeV.
We would, of course, obtain a similar shift with the more physical 
value of $a=0.6$ fm in Fig.~1a if it were possible to do this calculation.
We do not insist too much on this fine detail of the problem since, as
already noted, there are ambiguities stemming from the HIVAP calculation.
We have also ignored other possible coupling effects such as neutron-transfer
channels, though these will always have unfavourable $Q$ values for 
symmetric systems. Fig.~2b shows the derivative of $T_0$ with respect to 
the incident energy for the $[4,2]$ and $[4,4]$ calculations. 
It is well known that this gives the distribution of
barriers $D(E)$~\cite{dtde}, and it can be seen that there is little 
difference between the two distributions except for the presence of a
lower adiabatic barrier with very small weight (barely visible on this scale)
in the latter case. We can, therefore, conclude that the calculations
have essentially converged. This is reassuring since the need to introduce
higher phonon states might be somewhat dubious. We note that the
adiabatic barrier of our calculations is not the same as that of Quint et al. 
which has a weight of 1 and is supposed simply to reproduce 
$T_0$  at low $E$.

The calculations that we have presented show the importance of higher
phonon couplings not included in the CCFUS calculations of Ref.~\cite{quint}.
There is, however, another very important difference which introduces
new physics into the barrier distribution, and which we shall now elaborate.

\begin{figure}
\vspace{100mm}
\includegraphics{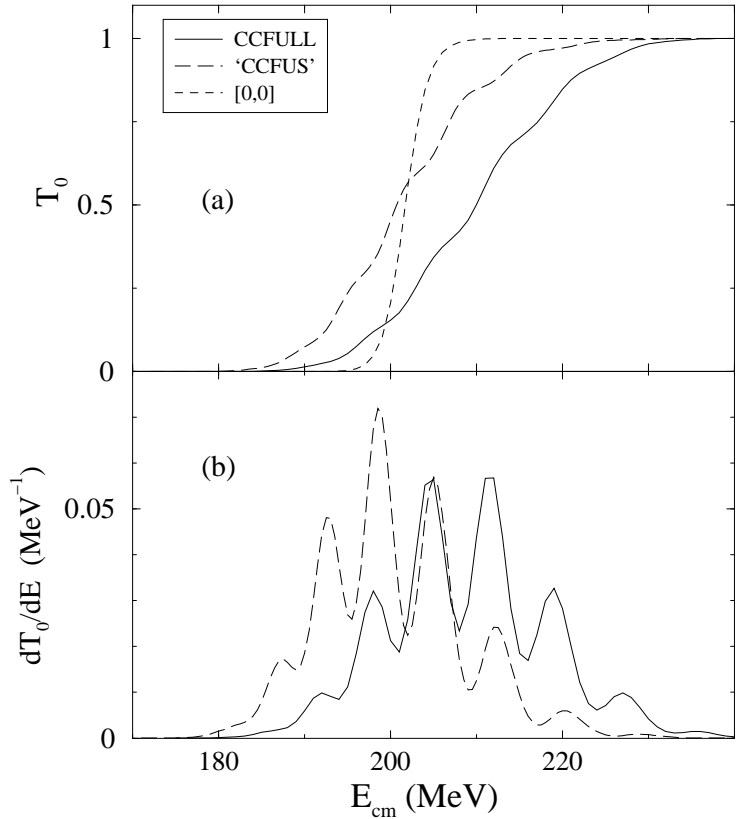}
\caption{The $[4,4]$ CCFULL calculation compared with a $[4,4]$
calculation in the spirit of CCFUS. See text for details. Note that
the latter calculation does not produce a shift of the $T_0=0.5$ point,
whereas the CCFULL calculation gives a shift of about 10 MeV due to the
higher weights of the high-$E$ barriers.
}
\label{fig3}
\end{figure}

In CCFUS, everything is essentially determined in the
barrier region, and the barrier heights and weights obtained through
the diagonalisation of the coupling matrix (including excitation energies)
at the barrier radius. This is probably a reasonable approximation
for the short-ranged nuclear field but will fail for heavy
systems where the Coulomb field plays an important role at large distances. 
In order to simulate a CCFUS-type model but still include all of the nuclear
$[4,4]$ couplings, we performed a calculation in which the Coulomb deformation
parameters were set to zero. However, this will also change the barrier
heights, since the deformed Coulomb field is not negligible at the barrier.
In order to correct for this, we renormalised the nuclear deformation
parameters (this is possible since the same
geometrical factors appear in both couplings). The results for the
relevant barrier distributions are shown in Fig.~3. One sees that the
barriers occur at almost exactly the same positions in the two calculations
but that in the complete calculation the weights are greatly shifted
towards the high-energy barriers, due to the Coulomb couplings at
large distances. In effect, the Coulomb field favours the linear 
superposition of states which lowers its own energy. Since it has the 
opposite sign from the nuclear field, this configuration is precisely
that which minimises the nuclear forces, that is, the one corresponding to 
the  {\em highest} barrier. In other words, the nuclei are polarised in the 
entrance channel to disfavour the lower barriers. The effect leads to an
overall shift of the barrier centroid of around 10 MeV, even though
the individual barrier positions remain unchanged. (The $T_0=0.5$
point of the CCFUS-type calculation is essentially unshifted.) 
Since $D_{\infty}$
in the present case is about 20 MeV, this gives the factor
of around 2 which was missing from $D_{\infty}$ in the calculations 
of Quint et al.

We believe that similar considerations apply to the work of Berdichevsky 
et al. 
\cite{norenberg} who used a single-particle model to approximately derive 
the barrier splittings but without doing a full calculation of 
the scattering. (They rather compared their spread of barriers with the 
$\Delta$ of Ref.~\cite{quint}.) Such a model may give a reasonable  
spread of barriers but it is important to have the relevant correlations
which render the nuclear states collective in order to get the correct
reaction dynamics and the correct shape of $T_0$.

We have obtained an excellent fit to the proposed shape of the capture
cross section with physically reasonable parameters. 
However, we should
now return to the question of what is the appropriate uncoupled
barrier height. Do our calculations retrieve the Bass barrier? The answer
to this question is no. Our uncoupled barrier is 201.7 MeV
and the Bass barrier 195.2 MeV. That is we still need an uncoupled barrier
6.5 MeV higher than $B_{\rm Bass}$ (previously 12.2 MeV~\cite{quint})
and we should ask why this is so.
There are various possible explanations for this including:
\begin{itemize}
\item The Bass potential contains a factor $R_1\,R_2/(R_1+R_2)$ which 
accounts for the curvature of the two nuclear surfaces. This 
factor is largest for symmetric systems and may simply over-estimate
the potential for such reactions, giving too low a barrier.
\item The Bass potential parameters are fitted to experimental data,
which necessarily contain all possible couplings. It is known that
high-lying phonon states shift the barrier centroid to lower energies
\cite{shifts}.
Thus the uncoupled barrier should probably be taken to be higher than the 
Bass barrier if one accounts for the couplings explicitly, as we do here.
\end{itemize}
We should not, however, forget the ambiguities in mapping from 
$\sigma_{\rm ER}$ to $\sigma_{\rm cap}$. These come
both from ambiguities in the statistical-model parameters and from 
the complete neglect of the QF process, and in this context it is 
interesting to look at other symmetric systems. Fig.~4 shows our fits
to the systems $^{90}$Zr + $^{90}$Zr~\cite{keller}
and $^{100}$Mo + $^{110}$Pd 
\cite{quint}. These will be discussed in detail elsewhere~\cite{graretal}.
Here we note simply that the barrier shift we require for 
$^{90}$Zr + $^{90}$Zr is 4.1 MeV, similar to that for $^{100}$Mo + $^{100}$Mo,
but for $^{100}$Mo + $^{110}$Pd we require a shift of 15 MeV
(previously 29.0 MeV), which
does not seem consistent with the other systems. However, it has been
pointed out~\cite{quint,keller} that if one performs the HIVAP calculations 
with a smaller shell-damping parameter (the energy range over which shell 
effects are smeared out) different solutions for the gaussian parameters 
(hence different $T_0$) are possible. The effects are relatively small for  
$^{90}$Zr + $^{90}$Zr and $^{100}$Mo + $^{100}$Mo, changing $\Delta$ 
rather little but moving $\bar B$ down to make our uncoupled barrier rather
closer to the Bass value. However, for the system $^{100}$Mo + $^{110}$Pd
(where the ratio $\sigma_{\rm ER}/\sigma_{\rm cap}$ is much smaller and 
$\sigma_{\rm QF}$ may also be more important) 
the effect is much larger, giving a shift down of around 8 MeV but still 
leaving the uncoupled barrier around 7 MeV higher than $B_{\rm Bass}$.

\begin{figure}
\vspace{100mm}
\includegraphics{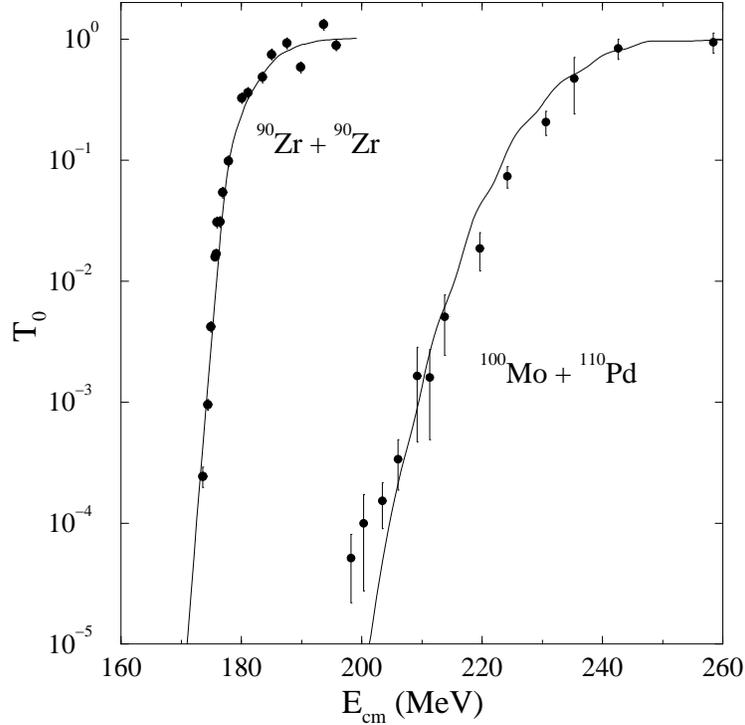}
\caption{CCFULL fits to two different systems. Large ambiguities exist
in the experimental curve for the heavier one. See text. 
}
\label{fig4}
\end{figure}

The ambiguities here are sufficiently important to merit
further experimental investigation. The most pertinent case is
$^{100}$Mo + $^{110}$Pd, and the ambiguity could be resolved by a 
direct measurement of $\sigma_{\rm cap}$ for this system,
as discussed at the beginning of this Letter. It might, however, be
simpler to exploit unitarity and obtain the capture barrier distribution
from the large-angle quasielastic flux scattered back from the 
Coulomb barriers~\cite{timmersqe,haginoqe,piasecki}. 

This work was supported by the Grant-in-Aid for Scientific Research,
Contract No. 16740139,
from the Japanese Ministry of Education, Culture, Sports, Science and
Technology.

\end{document}